\documentclass[]{aa}

\usepackage{graphicx}
\usepackage{color}
\usepackage{mathtools}
\usepackage{amsfonts}
\usepackage{amsmath}
\usepackage{amssymb}
\usepackage{bm}
\usepackage{float}
\usepackage{svg}
\usepackage{tabularx} 
\usepackage{subcaption}
\usepackage{hyperref}

\def\d{{\rm d}}
\def\obs{{\rm o}}
\def\rest{{\rm r}}

\def\bth{\bm{\theta}}
\def\bbeta{\bm{\beta}}
\def\Slim{S_{\rm lim}}
\def\Llim{L_{\rm lim}}

\begin{document}

\title{The kinematic cosmic dipole beyond Ellis and Baldwin}
\author{Albert Bonnefous\inst{1}\fnmsep\thanks{albert.bonnefous@iap.fr}}
\institute{Sorbonne Université, CNRS, Institut d'Astrophysique de Paris,  98bis Boulevard Arago, 75014 Paris, France}
\date{Received YY 20ZZ}

\abstract{The cosmic dipole anomaly—currently detected at a significance exceeding 5$\sigma$ in several independent surveys poses a significant challenge to the standard model of cosmology. The Ellis and Baldwin formula provides a theoretical link between the intrinsic dipole anisotropy in the sky distribution of extragalactic light sources and the observer's velocity relative to the cosmic rest frame, under the assumptions that the sources follow a power-law luminosity function and exhibit power-law spectral energy distributions. Even though for a monochromatic survey fitting a power law on the spectra at the flux limit is always sufficient, it fails for the case of sources with more complicated spectra in photometric surveys, such as galaxies in the visible and near-infrared, which can feature emission lines or breaks. In this work, we demonstrate that the Ellis and Baldwin formula can be generalized to arbitrary luminosity distributions and spectral profiles in particular for photometric surveys. We derived the corresponding expression for the effective spectral index and applied it to a sample of quasars observed in the W1 band of the CatWISE survey. We show that the anomalous cosmic dipole persists beyond the power-law assumption in this sample. These results provide a more general and robust framework to interpret measurements of the cosmic dipole in future photometric large-scale surveys.}


\maketitle
\nolinenumbers

\section{Introduction}

In their 1984 article \citep{ellis_expected_1984}, Ellis and Baldwin pointed out that our motion with respect to the cosmological rest frame (CRF) induces intrinsic anisotropy in the observed sky distribution of any class of objects in the Universe. In other words, the Universe does not appear perfectly homogeneous and isotropic to us. In the specific case of radio sources, whose spectra and flux distributions can be well approximated by power laws, they proposed a method to infer our velocity in the CRF by measuring this anisotropy, more precisely the dipole in the number of sources per unit solid angle. They derived what we hereafter refer to as the Ellis and Baldwin (EB) formula:
\begin{equation}
\bm{\mathcal{D}}_{\rm kin}=(2+x(1+\alpha))\bbeta\,,
\end{equation}
where $x$ and $\alpha$ denote the power-law indices of the flux distribution and source spectra, respectively, and $\bbeta$ is our velocity relative to the CRF, normalized by the speed of light. Ellis and Baldwin argued that estimating $\bbeta$ in this way should yield the same result as the velocity measured from the dipole of the cosmic microwave background (CMB), provided that the sources are sufficiently distant for intrinsic clustering anisotropies to be neglected and lie entirely beyond our local bulk flow, within a Friedmann-Lemaître-Robertson-Walker (FLRW) Universe.

At the time, however, applying this test to observational data was not feasible, because sufficiently large catalogues were not available to obtain a robust signal-to-noise ratio in the dipole measurement. The first detection of a kinematic dipole with this test dates back to 2002, with the NRAO VLA Sky Survey (NVSS), which included about 1.8 million sources \citep{blake_detection_2002}. The measured dipole amplitude was slightly higher than that expected from the CMB, although its low significance meant that it could still be compatible. Since then, other tests aligned with this result, which showed correct alignment but a higher than expected amplitude (see for example \citep{singal_large_2011,gibelyou_dipoles_2012,rubart_cosmic_2013,colin_high-redshift_2017}). In 2021, a robust result of about $5\sigma$ in statistical significance, was obtained using CatWISE quasar data \citep{secrest_test_2021}, and in 2022 it was shown that NVSS radio galaxies yield a similar result \citep{secrest_challenge_2022}. Despite being uncorrelated and affected by distinct systematics, both datasets yielded inferred velocities significantly higher than the CMB expectation. Within the $\Lambda$ cold dark matter ($\Lambda$CDM) framework, no explanation has yet been found. Various systematics were invoked as the reason for this anomaly, which have been shown to be mostly inconsequential (see {e.g.} \cite{darling_universe_2022}). It has also been shown that plausible origins, within the $\Lambda$CDM framework, such as weak lensing by local structures that could potentially provide a physical explanation of this dipole are not strong enough to fully reconcile both the matter and the radiation dipoles \citep{murray_effects_2022,bonnefous_weak_2026}.

These findings have enforced the validity of the cosmic dipole anomaly, and the EB formalism itself was also scrutinized. In particular, the expression for the kinematic dipole in number counts has been generalized to a formulation that incorporates the redshift evolution of the source description \citep{maartens_kinematic_2018}. These indices $x$ and $\alpha$ have also been linked to the evolution of the luminosity function \citep{dalang_kinematic_2022}. It has been shown that the original EB formula remains correct in the presence of source evolution and that for monochromatic surveys, it is necessary to evaluate the parameters $x$ and $\alpha$ at the flux limit \citep{von_hausegger_expected_2024}. A more detailed review of this test and its use over the years is available in \citep{secrest_colloquium_2025}.

\begin{figure}
    \centering
    \includegraphics[width=0.9\linewidth]{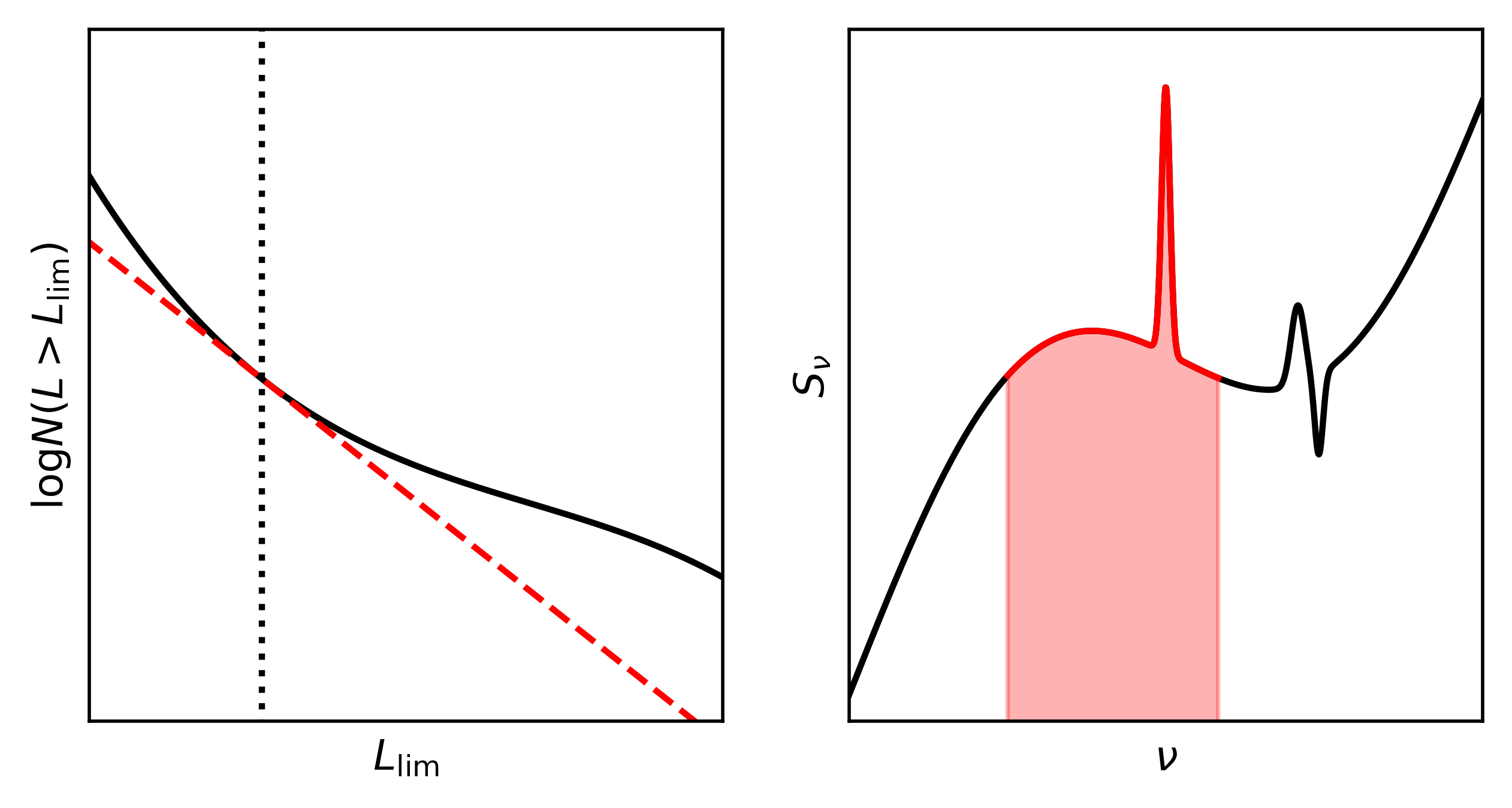}
    \caption{Left: Sketch of an integral number count with a power-law fit (dashed red line) at an arbitrary flux or luminosity limit. Right: Arbitrary spectrum of a source at the flux or luminosity limit, where the red part corresponds to the part of the spectrum integrated in a given band $X$. In the most general case survey, it is always possible to fit a power law at the flux or luminosity limit. However, here the spectra are not a power law, so for a photometric survey the spectral index $\alpha$ cannot be defined as $S_\nu\propto\nu^{-\alpha}$ over the band.}
    \label{fig:sketch}
\end{figure}

However, the case of sources, which do not exhibit power-law spectra in photometric survey, has not been treated yet. Solving this issue is crucial for applying this test to upcoming large-scale surveys such as LSST \citep{ivezic_lsst_2019}, Euclid \citep{euclid_collaboration_euclid_2025}, and SPHEREx \citep{crill_spherex_2020}. For example, the spectra of galaxies in the visible and near-infrared can feature emission lines, bumps, or even more complicated features such as a Lyman-$\alpha$ forest (see Fig.~\ref{fig:sketch}). This is the topic of our study. Note that a full generalization of the Ellis and Baldwin formula has also been recently proposed in \citep{takeuchi_general_2026}, using a different formalism involving selection functions.

In this paper, we first briefly review the formalism underlying the EB formula in Sect. \ref{sec:EB_formula}. We then extend it in Sect. \ref{sec:kin_dipole_general} to the general case of a source population without assuming any specific spectral profile or luminosity distribution. We consider two distinct survey types: monochromatic (measuring flux density, as is typical of radio surveys) and photometric (the flux integrated with a pass-band filter). Finally, in Sect. \ref{sec:alpha_quasar}, we apply the resulting effective spectral index in photometric surveys to quasars and compare it to the value obtained using the approach of \citep{secrest_test_2021}.

\section{The Ellis and Baldwin formula}
\label{sec:EB_formula}
\begin{figure}
    \centering
    \includegraphics[scale=0.65]{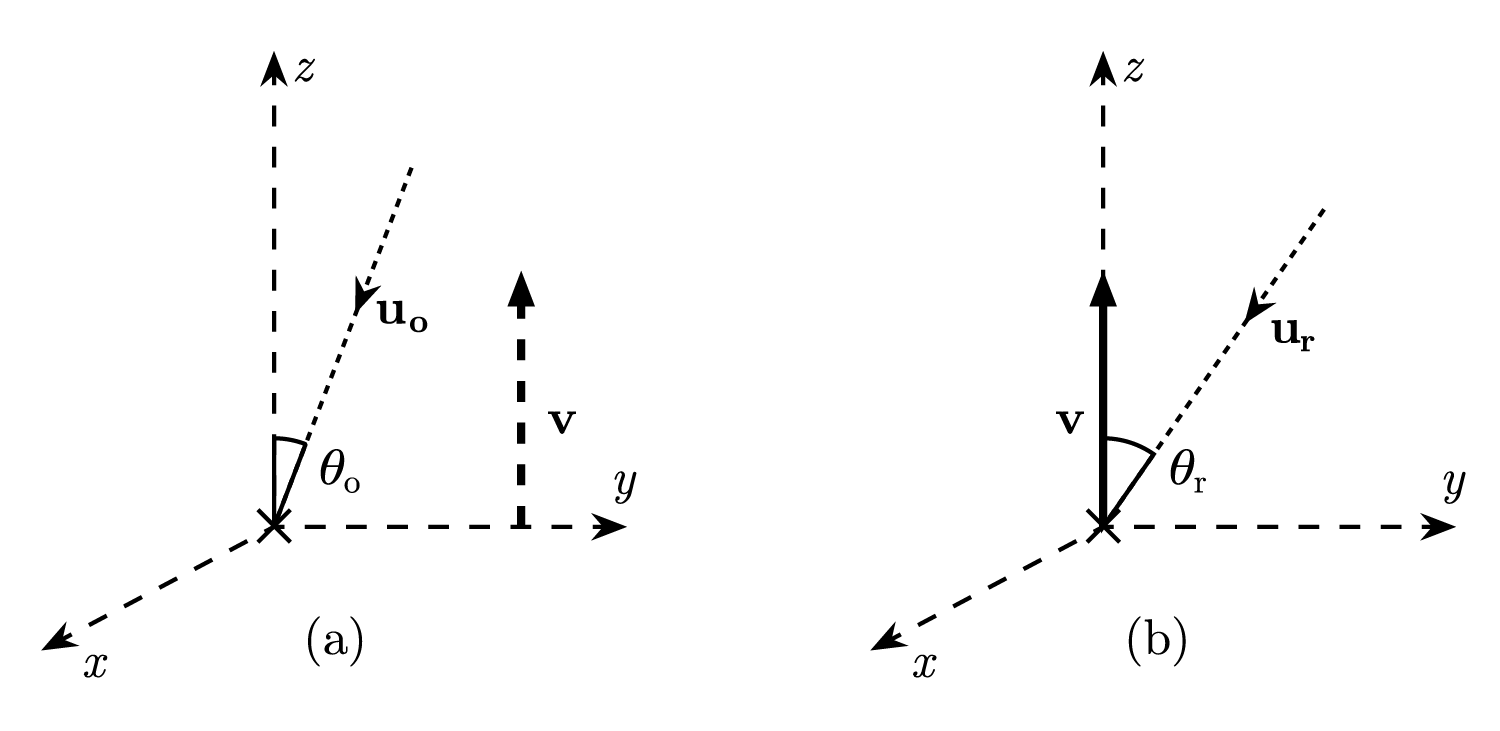}
    \caption{(a) Frame moving with the observer. (b) Rest frame, in which the observer has a velocity $\mathbf{v}$. For simplicity, we assume that the $z$ axis is parallel to the observer's movement.}
    \label{fig:frame}
\end{figure}

Consider a moving observer with respect to a rest frame, such as that presented in Fig.~\ref{fig:frame}. Any quantity $a$ measured in the frame of the observer is written as $a_\obs$, whereas the same quantity in the rest frame is $a_\rest$. Two different effects impact the observations, the relativistic aberrations, and the Doppler boosting.

\subsection{Relativistic aberration}
The relativistic aberration of light is the deformation of an observer's field of view when moving relative to what it observes; the apparent position of a body shifts towards the direction of its movement. Suppose that we observe an object at a position $\bth_\obs=(\theta_\obs,\phi_\obs)$ in the sky. In the rest frame, the angle $\phi_\rest$ does not change, and $\theta_\rest$ transforms such that
\begin{equation}
    \sin{\theta_\rest} =\frac{\sin{\theta_\obs}}{\gamma(1-\beta\cos{\theta_\obs})}\,,
\end{equation}
where $\gamma=1/(1-\beta^2)^{1/2}$. With $\d\Omega =\sin{\theta}\d\theta\d\phi$, we obtain the distortion of the solid-angle relation
\begin{equation}\label{eq:aberration}
    \d\Omega_\obs=\delta(\bth_\obs)^{-2}\d\Omega_\rest\,,
\end{equation}
and introduce the notation
\[
\delta(\bth)=\frac{1}{\gamma(1-\bbeta\cdot\bth)}\approx 1+\bbeta\cdot\bth+\mathcal{O}(\beta^2)\,.
\]
Also, for the sake of simplicity, the $\mathcal{O}(\beta^2)$ is omitted in every approximation that follows.

\subsection{Relativistic Doppler effect}
The Doppler boosting effect describes the light coming from a distant object as brighter and bluer in the observer's direction of movement. Between the two frames, the frequency of a given photon shifts as
\begin{equation}
    \nu_\obs=\delta(\bth_\obs)\nu_\rest\,.
\end{equation}
The flux density $S=\frac{\d E}{\d\nu \d t}$ also transforms between frames as
\begin{equation}
    S_\obs(\nu_\obs)=\delta(\bth_\obs)S_\rest(\nu_\rest)\,.
\end{equation}

\subsection{Ellis and Baldwin formula with power laws}\label{subsec:EB_formula}
Now, let us assume that the rest frame is the CRF. If the observer does not move with respect to this frame, it would observe a completely uniform background of $N$ objects, assuming that they are sufficiently distant for local inhomogeneities to be negligible \citep{gibelyou_dipoles_2012}. These objects can be, for example, galaxies or quasars, but we assume only one type of object in our data. Ellis and Baldwin \citep{ellis_expected_1984} make the following assumptions. First, every object's spectral flux density $S$ has a simple power-law dependence on the frequency $\nu$:
\begin{equation}\label{eq:power_law_alpha}
    S\propto \nu^{-\alpha}\,.
\end{equation}
Secondly, the objects have a power-law distribution in the number of objects $N$ above a certain flux $\Slim$:
\begin{equation}\label{eq:power_law_x}
    N(>\Slim)\propto \Slim^{-x}\,.
\end{equation}
These two indices $x$ and $\alpha$ completely depend on the type of object we observe, their redshift distribution, and the chosen flux cut. In the observer's frame, the flux received from one source is boosted by the Doppler effect, and using the power law \eqref{eq:power_law_alpha}, the flux received from this object at a fixed frequency $\nu_\obs$ is
\begin{equation}
    S_\obs(\nu_\obs)=\delta(\bth_\obs)^{1+\alpha}S_\rest(\nu_\obs)\,.
\end{equation}
The number of objects, in a given direction $\bth_\obs$, then becomes
\begin{equation}\label{eq:dN_obs_rest}
\begin{split}
    \d N_\obs(>\Slim,\bth_\obs)&=\d N_\rest(>\delta(\bth_\obs)^{-(1+\alpha)}\Slim,\bth_\obs)\\
    &=\delta(\bth_\obs)^{x(1+\alpha)}\d N_\rest(>\Slim,\bth_\obs)\,.
\end{split}
\end{equation}
Note that here, $\Slim$ is fixed. Rather, it is the objects' flux density that changes—as if the flux density limit depended on the frame and thus on direction. Finally, accounting for the aberration of solid angles \eqref{eq:aberration}, the number count of objects per solid angle $N_\Omega=\frac{\d N}{\d \Omega}$ in the observer frame becomes
\begin{equation}\label{eq:number_count}
\begin{split}
    N_{\Omega,\obs}(>\Slim,\bth_\obs)&=\frac{\d N_\obs}{\d\Omega_\obs}(>\Slim,\bth_\obs)\\
    &=\delta(\bth_\obs)^{2+x(1+\alpha)}N_{\Omega,\rest}(>\Slim)\,.
\end{split}
\end{equation}
If we integrate this formula on the whole sky, the overall number of observed sources in the sky does not change with the frame at order $\beta$ (see further details in Appendix). Therefore, we write 
\[N_{\Omega,\rest}(>\Slim)\approx\bar{N}_{\Omega}(>\Slim)\approx N(>\Slim)/4\pi\,.
\]
Note that if we want to account for any other non-kinematic cosmological perturbations in the number count, we can do so by including them in this $\bar{N}_{\Omega}$ \citep{nadolny_new_2021}. Now, if we take the CMB as a reference frame, we have $v=369.82 \pm 0.11\,{\rm km.s^{-1}}$ \citep{planck_collaboration_planck_2020}. Therefore $\beta\ll1$, and the number count \eqref{eq:number_count} can be expressed purely with observed quantities via
\begin{equation}
    N_{\Omega,\obs}(>\Slim,\bth_\obs)\approx\bar{N}_\Omega(1+\bm{\mathcal{D}}_{\rm kin}\cdot\bth_\obs)
\end{equation}
\begin{equation}\label{eq:EB_formula}
    \bm{\mathcal{D}}_{\rm kin}=(2+x(1+\alpha))\bbeta.
\end{equation}
This is the form derived by Ellis and Baldwin \citep{ellis_expected_1984}. We have seen that this expression assumes that the spectrum and luminosity function of these objects are power laws, and—most importantly—that a CRF exists in which these objects form a uniform background. However, since aberration and Doppler boosting are purely local relativistic effects, this formula is completely independent of the cosmological model, provided that the cosmological principle holds. This formula also requires no knowledge of the objects' redshift, only their location in the sky.

\section{Kinematic dipole without any assumption on the spectrum and number count}
\label{sec:kin_dipole_general}
Here, we derive a dipole expression, without making the power-law hypothesis for the cumulative number count and spectrum of our sources. We assume that $\beta\ll1$ and that all the sources exhibit the same spectrum, or at least the same spectral profile. In other words $S_i\propto s(\nu)$ for every object $i$ in our survey, and $s(\nu)$ is any function.

\subsection{Monochromatic survey}
First, we consider a survey at a single wavelength $\nu_\obs$. At this wavelength, the spectral flux density $S$ transforms as
\begin{equation}
\begin{split}
    S_\obs(\nu_\obs)&=\delta(\bth_\obs)S_\rest(\nu_\obs\delta(\bth_\obs)^{-1})\\
    &\approx(1+\bbeta\cdot\bth_\obs)\left(S_\rest(\nu_\obs)-\nu_\obs\frac{\partial S_\rest}{\partial\nu_\obs}\bbeta\cdot\bth_\obs\right)\,.
\end{split}
\end{equation}
We obtain $S_\obs(\bth_\obs,\nu_\obs)=S_\rest(\nu_\obs)+\delta S(\bth_\obs,\nu_\obs)$, where $\delta S(\bth_\obs,\nu_\obs)$ is defined as
\begin{equation}
\begin{split}
    \delta S(\bth_\obs,\nu_\obs)&\approx\left(S_\rest(\nu_\obs)-\nu_\obs\frac{\partial S_\rest}{\partial\nu_\obs}\right)\bbeta\cdot\bth_\obs\\
    &\approx S(\nu)\left(1-\frac{\partial\log S}{\partial\log\nu}\right)\bbeta\cdot\bth_\obs\,.
\end{split}
\end{equation}
Note that here, every quantity in the rest frame can be expressed in the observer frame since the difference is in $\mathcal{O}(\beta)$ and becomes $\mathcal{O}(\beta^2)$ after multiplication by $\bbeta\cdot\bth_\obs$. Therefore, for simplicity in the following we drop the indices indicating the frame for every term proportional to $\bbeta$. We then expend the number count around the limiting flux $\Slim$. Since $\delta S$ is of order $\mathcal{O}(\beta)$, we have
\begin{equation}
\begin{split}
    &N_{\Omega,\obs}(>\Slim,\bth_\obs)=\delta(\bth_\obs)^2N_{\Omega,\rest}(>\Slim-\delta\Slim,\bth_\rest)\\
    &\approx(1+2\bbeta\cdot\bth_\obs)\left(\bar{N}_\Omega(>\Slim)-\frac{\partial \bar{N}_{\Omega}}{\partial\Slim}\delta \Slim\right)\\
    &\approx\bar{N}_\Omega
    \left(1+\left(2-\frac{\partial\log N}{\partial\log\Slim}\left(1-\frac{\partial\log \Slim}{\partial\log\nu}\right)\right)\bbeta\cdot\bth_\obs\right)\,.
\end{split}
\end{equation}
As before, in the rest frame, $N_{\Omega,\rest}(>S_{\rm lim,\rest},\bth_\rest)\approx N(>\Slim)/4\pi$, and the kinematic dipole in the observer frame is
\begin{equation}
    \bm{\mathcal{D}}_{\rm kin}=\left(2-\frac{\partial\log N}{\partial\log\Slim}\left(1-\frac{\partial\log \Slim}{\partial\log\nu}\right)\right)\bbeta\,.
\end{equation}
This expression is strictly equivalent to the EB formula \eqref{eq:EB_formula}, where we take the effective coefficient given by
\begin{equation}
    x_{\rm eff}=-\frac{\partial\log N}{\partial\log\Slim}
\end{equation}
at the limiting flux density $\Slim$. This is expected since only objects with a flux density close to the limit appear or disappear from the number count with Doppler boosting:
\begin{equation}
    \alpha_{\rm eff}=-\frac{\partial\log \Slim}{\partial\log\nu}
\end{equation}
at the frequency $\nu_\obs$. These results are similar to those obtained in \citep{von_hausegger_expected_2024}. Here since we consider all of our sources to have the same spectral profile, this $\alpha_{\rm eff}$ need not be expressed at $\Slim$ specifically. In the general case, however, this spectral index has to be taken for the sources at the flux limit, which are those susceptible to appearing or disappearing with Doppler boosting. In particular, it is notable that when the observed wavelength is close to a strong emission line and every source has the same redshift, this $\alpha_{\rm eff}$ could become very large—either negatively or positively—depending on the position of this emission line with $\nu_\obs$. This could lead to a sudden jump in the number count dipole. However, an important property of the spectral index is lost in this case: it is no longer redshift-independent. If we account for the fact that the spectra of all sources are shifted according to their individual redshifts, $S_i(\nu)\rightarrow S_i^{\rm int}((1+z_i)\nu)$, where $S_i^{\rm int}$ is the intrinsic spectra of the sources, the $\alpha_{\rm eff}$ is also modified. However, this does not affect the power law, as the redshift does not impact the spectrum profile, and it remains a power law.

\subsection{Photometric survey}
In practice, cosmological surveys often do not scan the sky at a monochromatic frequency $\nu_\obs$ but give the luminosity (or equivalently, the magnitude) of every object in a given band $X$. This luminosity can be expressed using the transmission function of the filter $T_X(\nu)$, here in terms of received energy:
\begin{equation}
    L=\int\d\nu\,T_X(\nu)S(\nu)\,.
\end{equation}
Note that here we neglected the transmission function of the atmosphere itself for ground-based telescopes, but the effect of atmosphere can be included in this filter transmission function. Suppose that we observe a particular object in the direction $\bth_\obs$ whose spectral flux density is $S_\obs(\nu_\obs)$, its luminosity transforms in the observer frame as
\begin{equation}
\begin{split}
    &L_\obs(\bth_\obs)=\int\d\nu_\obs\,T_X(\nu_\obs)S_\obs(\nu_\obs)\\
    &=\delta(\bth_\obs)^2\int\d\nu_\rest\,T_X(\delta(\bth_\obs)\nu_\rest)S_\rest(\nu_\rest)\\
    &\approx(1+2\bbeta\cdot\bth_\obs)\int\d\nu_\rest\,\left(T_X(\nu_\rest)+\bbeta\cdot\bth_\obs\nu_\rest\frac{\partial T_X}{\partial\nu_\rest}\right)S_\rest(\nu_\rest)\,.
\end{split}
\end{equation}
We obtain $L_\obs(\bth_\obs)=L_\rest+\delta L(\bth_\obs)$, where $\delta L(\bth_\obs)$ is defined as
\begin{equation}
\begin{split}
    \delta L(\bth_\obs)&\approx\left(2L+\int\d\nu\,\nu\frac{\partial T_X}{\partial\nu}S(\nu)\right)\bbeta\cdot\bth_\obs\\
    &\approx\left(L-\int\d\nu\,T_X(\nu)\nu\frac{\partial S}{\partial\nu}\right)\bbeta\cdot\bth_\obs\,.
\end{split}
\end{equation}
Note that we set $T_X(\nu)=T_X'(\nu)=0$ outside a particular frequency interval. For the same reason as $\delta S$ in the monochromatic survey, every quantity here can be expressed either in the rest frame or in the observer frame. The number count becomes
\begin{equation}
\begin{split}
    &N_{\Omega,\obs}(>\Llim,\bth_\obs)=\delta(\bth_\obs)^2N_{\Omega,\rest}(>\Llim-\delta\Llim,\bth_\rest)\\
    &\approx(1+2\bbeta\cdot\bth_\obs)\left(\bar{N}_{\Omega}(>\Llim)-\frac{\partial\bar{N}_{\Omega}}{\partial\Llim}\delta \Llim\right)\\
    &\approx\bar{N}_\Omega+\\
    &\bar{N}_\Omega\left(2-\frac{\partial\log N}{\partial\log\Llim}\left(1-\frac{\int\d\nu\,T_X(\nu)\nu\frac{\partial\Slim}{\partial\nu}}{\Llim}\right)\right)\bbeta\cdot\bth_\obs\,,
\end{split}
\end{equation}
where $\Slim(\nu)$ here is such as that its integrated luminosity is $\Llim = \int\d\nu\,T_X(\nu)\Slim(\nu)$. As before, we can define effective coefficient $x_{\rm eff}$ and $\alpha_{\rm eff}$ as
\begin{equation}
    x_{\rm eff}=-\frac{\partial\log N}{\partial\log\Llim}=+2.5\frac{\Phi(m_{\rm lim})}{N}\,,
\end{equation}
with $\Phi(m)=\frac{\d N}{\d m}$ being the luminosity function of the survey, $m_{\rm lim}$ the magnitude corresponding to $\Llim$, and
\begin{equation}\label{eq:alpha_eff}
    \alpha_{\rm eff}=-\frac{\int\d\nu\,T_X(\nu)\nu\frac{\partial \Slim}{\partial\nu}}{\int\d\nu\,T_X(\nu)\Slim(\nu)}=1+\frac{\int\d\nu\,\frac{\partial T_X(\nu)}{\partial\nu}\nu \Slim(\nu)}{\int\d\nu\,T_X(\nu)\Slim(\nu)}\,.
\end{equation}
This effective index characterizes how luminosity evolves with a spectral shift, similar to the previous spectral index. The previous remark about needing this spectral index at the flux limit in the general case remains valid here, even though it can be simplified under our assumption that all sources share the same spectral profile. Note that with $T_X(\nu)=\nu_\obs\delta(\nu-\nu_\obs)$ for a given wavelength $\nu_\obs$, we obtain the same coefficients as for the monochromatic survey, as expected. If we also assume $S\propto s(\nu)=\nu^{-\alpha}$, then $\alpha_{\rm eff}=\alpha$, regardless of the filter $T_X(\nu)$. We showed that the EB formula \eqref{eq:EB_formula} remains valid for any spectrum profile and luminosity function in the photometric survey, provided that the $x$ and $\alpha$ coefficients are defined with more precision. However, as before, we lose the redshift independence of this spectral index, and a single intrinsic galactic spectrum exhibits a different $\alpha_{\rm eff}$ depending on how much it has been redshifted.

Now, it is interesting to estimate the impact of an emission line on this index $\alpha_{\rm eff}$. Suppose that we observe a spectra $S_{\rm tot}(\nu) = S(\nu)+\varphi_{e.l}(\nu)$, where $\varphi_{e.l}(\nu)$ is an emission line, centred around a frequency $\nu_{e.l}$ within the band $X$. We can make a supposition on the shape of this emission line:
\begin{itemize}
    \item It is small enough so that it does not significantly impact the overall luminosity in the band, {i.e.} $\int\d\nu\,T_X(\nu)S_{\rm tot}(\nu)\approx\int\d\nu\,T_X(\nu)S(\nu)$.
    \item It is also thin enough to neglect the variation of $T_X(\nu)\nu$ on the domain where $\varphi_{e.l}\neq 0$.
\end{itemize}
With these two reasonable assumptions, it can be easily shown that this emission line $\varphi_{e.l}$ does not impact the effective index $\alpha_{\rm eff}$, as the impact of its derivative on one side is compensated by the other during integration.

\section{Effective spectral index for quasars}
\label{sec:alpha_quasar}
Now that we have an expression for the general effective coefficient for any spectral profile, we apply this new expression of $\alpha_{\rm eff}$ to real data as a proof of concept. This allows us to probe the limitations of this formula and check its consistency. In particular, we want to see whether the method for obtaining the $\alpha$ coefficient used in \citep{secrest_test_2021} (denoted as $\alpha_{\rm mag}$ in the following) yields the same results as ours here. Using the W1 and W2 bands of the CatWISE survey, this method compares the W1-W2 magnitude difference with a table of the same quantity calculated for synthetic pure power-law spectra with different $\alpha$ coefficients. We expect these three methods to yield similar results, as the quasar's spectra in this band should be well described by power laws.

To obtain $\alpha_{\rm eff}$ coefficients, we needed quasar spectra from the W1 band, which expands from $2.7$ to $3.9{\rm~\mu m}$ \citep{eisenhardt_catwise_2020}. We used the AKARI QSONG catalogue \citep{myungshin_akari_2017}, which is a compilation of $753$ quasar spectra taken between $2.5$ and $5{\rm~\mu m}$. However, to ensure good spectral quality, we only selected spectra with mean spectral flux densities above $5{\rm~mJy}$ and positive $\alpha_{\rm fit}$ values. This yields 41 individual quasars from which we extracted their corresponding W1- and W2-band magnitudes from the CatWISE2020 catalogue \citep{marocco_catwise2020_2021}. We also used the transmission filter $T_{\rm W1}(\nu)$, given by CatWISE \footnote{The transmission in response per photon is given here \url{https://www.astro.ucla.edu/~wright/WISE/passbands.html}}.

We applied Eq. \eqref{eq:alpha_eff} to these spectra to obtain the effective spectral index $\alpha_{\rm eff}$. We used the {\texttt Python} package {\texttt uncertainty} to estimate the error bars \footnote{See \url{https://pythonhosted.org/uncertainties/}}. We note that a naive integration of \eqref{eq:alpha_eff} can lead to huge error bars in the result: the spectral gradient is often smaller than its uncertainty here. However, the gradient is not a set of independent variables. Rather, it is constrained by the value of its sum; therefore, its uncertainty can be easily over-estimated. Error propagation must be carefully handled, and it may be more appropriate to use the last term of Eq. \eqref{eq:alpha_eff}, which avoids this gradient. Thus, we can infer the corresponding spectral index $\alpha_{\rm mag}$ using the CatWISE magnitude. Lastly, we also fit the spectra within the W1 band using a power law to obtain a final spectral index $\alpha_{\rm fit}$.

\begin{figure}
    \includegraphics[width=0.9\linewidth]{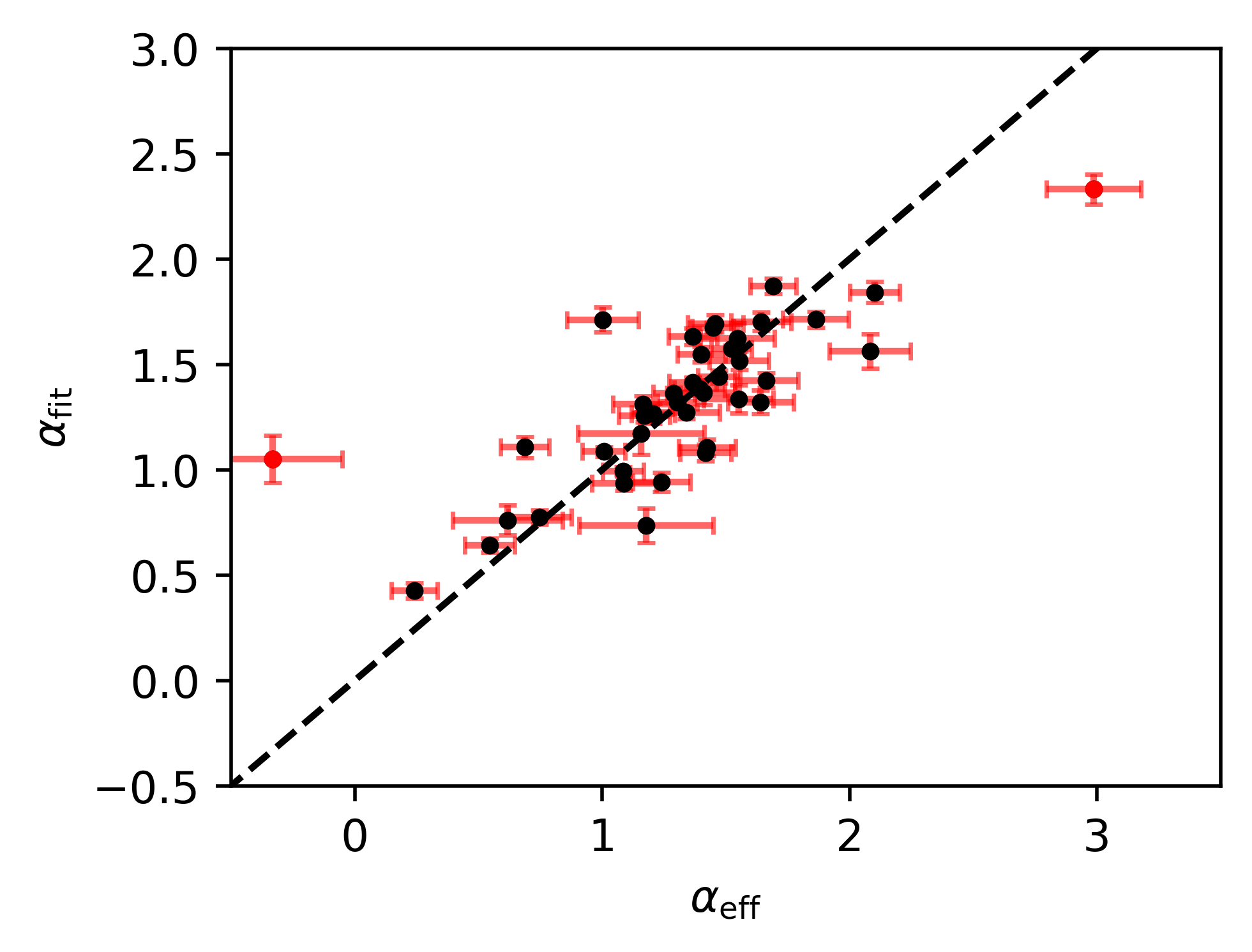}
    \includegraphics[width=0.9\linewidth]{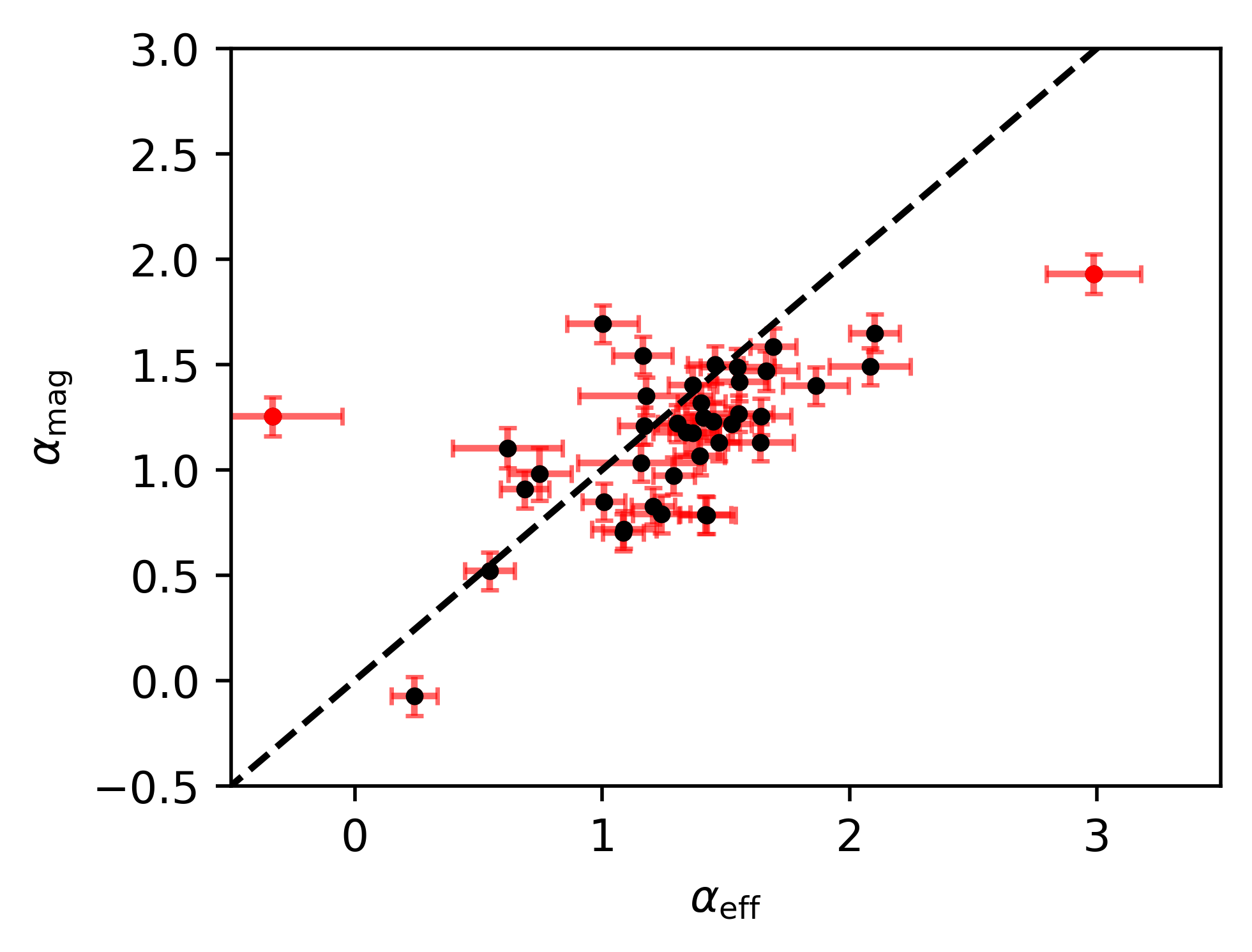}
    \includegraphics[width=0.9\linewidth]{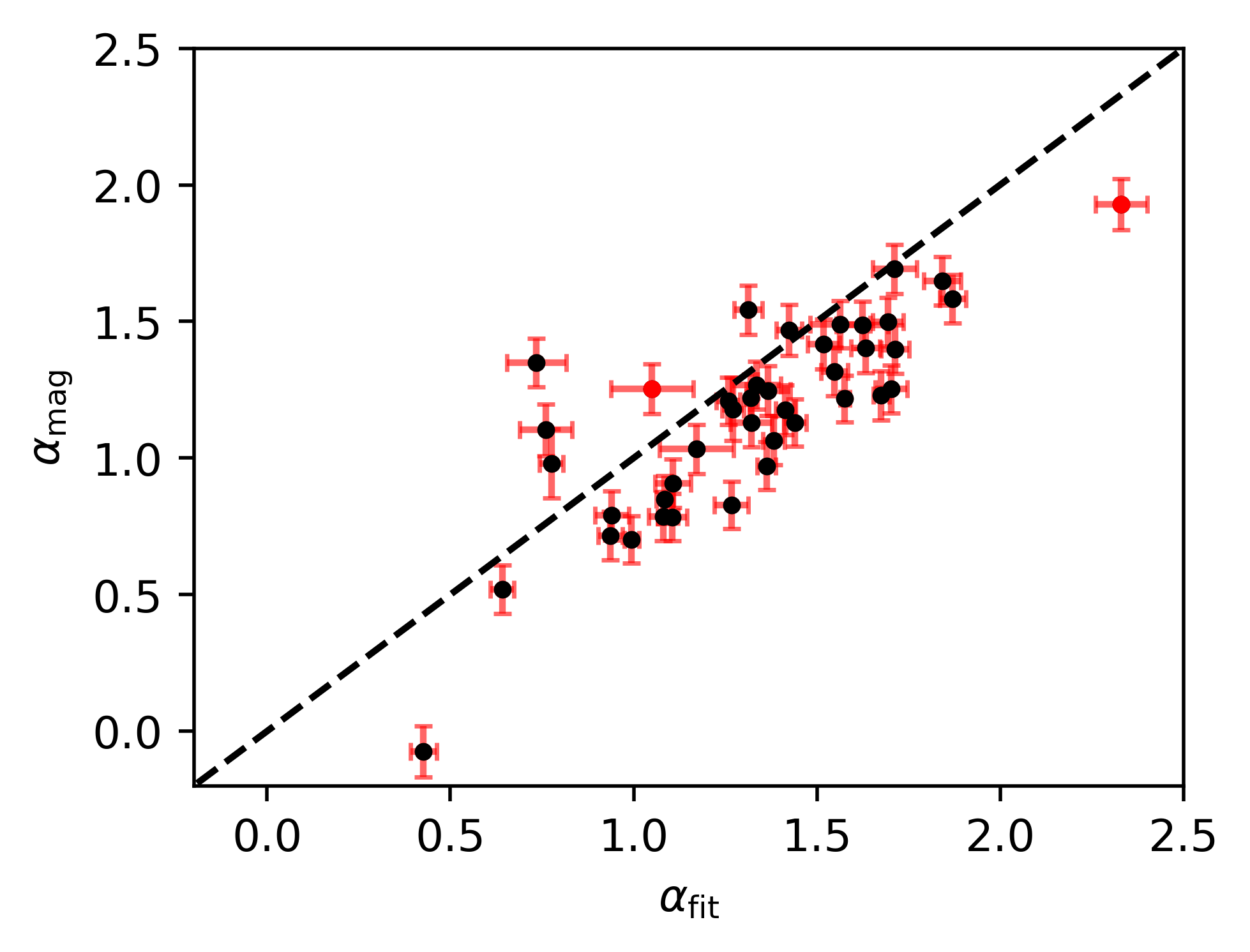}
    \caption{Comparison of the spectral index $\alpha$ obtained with different methods for 41 quasar spectra. The dashed black line corresponds to the identity. The two red dots with a high deviation between their calculated $\alpha_{\rm eff}$ and the two other methods correspond to the two quasars with the lowest mean spectral flux densities.}
    \label{fig:alpha_comparison}
\end{figure}

\begin{figure}
    \centering
    \includegraphics[width=0.9\linewidth]{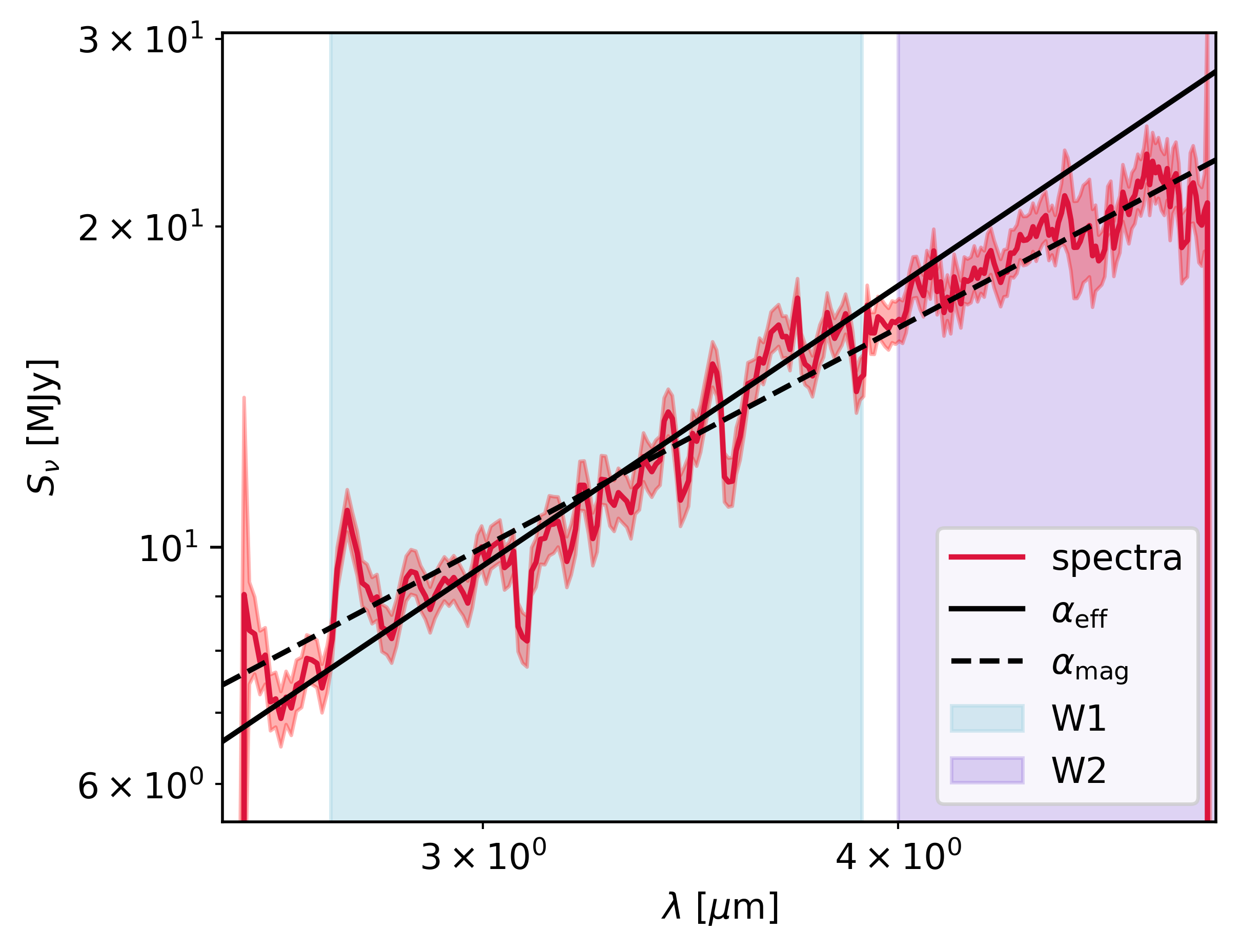}
    \caption{Example AKARI spectrum of quasar PG2112+059 at galactic coordinates $(272.101907, 64.128918)$, showing spectral indices $\alpha_{\rm mag}$ (obtained using the ${\rm W1-W2}$ method) and $\alpha_{\rm eff}$ (obtained from direct spectral integration via Eq. \eqref{eq:alpha_eff}). These spectral indices are plotted as pure power laws, $S_\nu\propto\nu^{-\alpha}$.}
    \label{fig:spectra_akari}
\end{figure}

We compare the different spectral indices obtained with these three methods in Fig.~\ref{fig:alpha_comparison}. We can see that the uncertainty when calculating $\alpha_{\rm eff}$ is quite large. This is explained by the large uncertainty of the spectra themselves. It is notable that the two points with the greatest deviation between our $\alpha_{\rm eff}$ and that from the other two methods correspond to the two spectra with the lowest spectral flux densities. This emphasizes the need for good spectral quality for precisely calculating $\alpha_{\rm eff}$. For all these quasars, we obtain
\[
\begin{split}
    \langle\alpha_{\rm eff}-\alpha_{\rm fit}\rangle & = -0.01 {\rm\ with\ } &\ \sigma = 0.33\,,\\
    \langle\alpha_{\rm eff}-\alpha_{\rm mag}\rangle & = 0.15 &\ \sigma = 0.42\,, \\
    \langle\alpha_{\rm fit}-\alpha_{\rm mag}\rangle & = 0.16 &\ \sigma = 0.22\,.\\
\end{split}
\]
We lack the statistical significance to make any definitive claim; however, nothing leads us to think that these three methods yield significantly different results for quasars in this ${\rm W1}$ band. At most, it seems that the $\alpha_{\rm mag}$ values are slightly smaller than the other spectral indices. If this method underestimates $\alpha$ by $0.16$, and the mean-measured $\alpha$ value for the quasars in \citep{secrest_test_2021} is $1.26$, then the quantity $2+x(1+\alpha)$ with $x=1.89$ would at most be underestimated by about $4\%$. This discrepancy might be explained by a slight difference in slope between the ${\rm W1}$ and ${\rm W2}$ bands. However, further investigation is necessary to obtain conclusive evidence. An example of a quasar spectrum from AKARI with its corresponding $\alpha_{\rm eff}$ and $\alpha_{\rm mag}$ is shown in Fig.~\ref{fig:spectra_akari}.

\section{Conclusion}
In this paper, we have demonstrated that the Ellis and Baldwin formula can be generalized to any class of light sources, regardless of their spectral shape or luminosity distribution, provided that the coefficients $x$ and $\alpha$ are appropriately redefined. We also showed that the impact of an emission line on the effective spectral index is negligible under the condition that this line is thin and small enough. This result is essential for addressing the cosmic dipole anomaly, as it shows that sources, such as galaxies with more complex spectral profiles than power laws in the near-infrared or visible light, can in principle be used to measure our velocity relative to the CRF with photometric surveys. Upcoming large-scale photometric surveys such as LSST \citep{ivezic_lsst_2019} or Euclid \citep{euclid_collaboration_euclid_2025} will provide us with the opportunity to apply the Ellis and Baldwin test to such data.

We also showed that, in the specific case of quasars, when using spectra from AKARI, the resulting effective coefficient $\alpha$ within the CatWISE ${\rm W1}$ band is generally consistent with the value obtained by fitting the spectrum with a pure power law. Moreover, we find that the procedure used in \citep{secrest_test_2021}, which relies only on ${\rm W1}$ and ${\rm W2}$ magnitudes, may only slightly underestimate the true spectral index by up to $\sim 0.16$. Still, this difference is not sufficient to significantly alter the conclusions of this work, which shows that the quasar dipole is a factor of $\sim2$ greater than the kinematic dipole expectation. Nevertheless, the limited number of available quasar spectra in this band prevents us from drawing definitive conclusions on this point. In the future, this issue may be clarified using spectroscopic libraries from missions such as SPHEREx \citep{crill_spherex_2020}.

In the general case, however, determining this effective spectral index $\alpha_{\rm eff}$ rigorously for a given dataset remains challenging. First, the spectral index is no longer independent of redshift: a given galaxy with fixed intrinsic spectra exhibits different values of $\alpha_{\rm eff}$ depending on its redshift. However, this effect can be understood statistically as a distribution of $\alpha$ values across sources, which adds to the intrinsic $\alpha$ distribution arising from the multiplicity of the sources' spectra themselves. Beyond this, another important issue is that most wide-area sky surveys are photometric, and the calculation of $\alpha_{\rm eff}$ requires a good quality of spectra within the range of the photometric band. Spectroscopic surveys generally have less objects, and this index may only be available for only a small subset of sources through cross-matching. Good quality spectra are generally only available for more luminous objects; therefore, measuring $\alpha_{\rm eff}$ for sources at the flux limit specifically would be complicated. This may introduce biases, which will have to be carefully handled. Future applications of this test will need to address these issues directly.

\begin{acknowledgement}
The author thanks Reza Ansari, Johann Cohen-Tanugi, Sebastian Von Hausegger, Roya Mohayaee,  Reiko Nakajima, Nathan Secrest for their important insights and  help.
\end{acknowledgement}

\bibliography{Beyond_EB.bib}
\newpage

\begin{appendix}
\section{Expansion of the Ellis and Baldwin formula at order $\beta^2$}
The coefficient $\delta(\bth_\obs)$ can be expanded as
\begin{equation}
    \delta(\bth_\obs)=1+\bbeta\cdot\bth_\obs+(\bbeta\cdot\bth_\obs)^2-\frac{1}{2}\bbeta^2+\mathcal{O}(\beta^3)\,.
\end{equation}
Therefore, we obtain
\begin{equation}
\begin{split}
    &\delta(\bth_\obs)^{2+x(1+\alpha)} = 1+(2+x(1+\alpha))\bbeta\cdot\bth_\obs\\
    &+(2+x(1+\alpha))\left(\frac{3+x(1+\alpha)}{2}(\bbeta\cdot\bth_\obs)^2-\frac{1}{2}\bbeta^2\right)+\mathcal{O}(\beta^3).
\end{split}
\end{equation}
Now, if we integrate Eq.~\eqref{eq:number_count} on the whole sky, the total number of sources observed and in the observer frame is
\begin{equation}
\begin{split}
    N_\obs &=\int\d\Omega_\obs\,N_{\Omega,\obs}(>\Slim,\bth_\obs)\\
    &=\int\d\Omega_\obs\,\delta(\bth_\obs)^{2+x(1+\alpha)}N_{\Omega,\rest}\\
     &\approx N_\rest\left(1+\bbeta^2\frac{x(1+\alpha)(2+x(1+\alpha))}{6}\right)\,.
\end{split}
\end{equation}
Here we omitted to write the $\mathcal{O}(\beta^3)$. It becomes apparent that the total number of sources actually changes depending on the frame, however it should be noted that this is a consequence of Doppler boosting, as setting $x=0$ removes the $\mathcal{O}(\beta^2)$ term. This is true at every order of $\beta$, and it can be shown quite easily that $\int\d\Omega\,\delta(\bth)^2=4\pi$. This is completely expected as aberration only changes the position of light sources in the sky. Then we can write
\begin{equation}
    N_{\Omega,\rest}(>\Slim)\approx\frac{N_\obs}{4\pi}\left(1-\bbeta^2\frac{x(1+\alpha)(2+x(1+\alpha))}{6}\right)\,.
\end{equation}
Finally, with $\bar{N}_{\Omega,\obs} = N_\obs/4\pi$, at order $\beta^2$, the number count becomes
\begin{equation}
\begin{split}
    &N_{\Omega,\obs}(>\Slim,\bth_\obs)\approx\bar{N}_{\Omega,\obs}(1+(2+x(1+\alpha))\bbeta\cdot\bth_\obs)\\
    &+\bar{N}_{\Omega,\obs}\frac{(2+x(1+\alpha))(3+x(1+\alpha))}{2}\left((\bbeta\cdot\bth_\obs)^2-\frac{1}{3}\bbeta^2\right)\,.
\end{split}
\end{equation}

\end{appendix}

\end{document}